\documentclass[aps,amsmath,amssymb,preprint,showpacs,superscriptaddress,endfloats*,pra]{revtex4}

\usepackage{mathrsfs}
\usepackage{leftidx}
\usepackage{graphicx, subfigure}  
\usepackage{epstopdf}
\usepackage{pgfplotstable, pgfplots} 
\usepackage{dcolumn}   
\usepackage{bm}        
\usepackage{braket}
\usepackage{comment} 
\usepackage{amssymb}   
\usepackage{enumerate}
\usepackage{amsmath}
\usepackage{float}
\usepackage[colorlinks,
            linkcolor=red,
            anchorcolor=blue,
			urlcolor=blue,
            citecolor=green]{hyperref}

\begin{document}
\title{16-qubit IBM universal quantum computer can be fully entangled}

\author{Yuanhao Wang}
\affiliation{Center for Quantum Information, Institute for Interdisciplinary Information Sciences, Tsinghua University, Beijing 100084, China}

\author{Ying Li}
\affiliation{Graduate School of China Academy of Engineering Physics, Beijing 100193, China}

\author{Zhang-qi Yin}
\email{yinzhangqi@tsinghua.edu.cn}
\affiliation{Center for Quantum Information, Institute for Interdisciplinary Information Sciences, Tsinghua University, Beijing 100084, China}

\author{Bei Zeng}
\affiliation{Department of Mathematics and Statistics, University of Guelph, Guelph N1G 2W1, Ontario, Canada}
\affiliation{Institute for Quantum Computing and Department of Physics and Astronomy, University  of  Waterloo, Waterloo N2L 3G1, Ontario, Canada}

 \begin{abstract}
Entanglement is an important evidence that a quantum device can potentially solve problems intractable for classical computers. In this paper, we prepare connected graph states involving 8 to 16 qubits on {\it ibmqx5}, a 16-qubit superconducting quantum processor accessible via IBM cloud, using low-depth circuits. We demonstrate that the prepared state is fully entangled, i.e. the state is inseparable with respect to any fixed partition.
\end{abstract}
\maketitle

\section{Introduction}
Quantum computation has been an active research topic since middle 90s with the invention of the Shor's algorithm and many other important discoveries such as quantum error correction~\cite{nielsen2002quantum}. For the last two decades, physical implementations of quantum computation have achieved significant progress. The fidelity of single and two-qubit gates exceeds 99\%, reaching the threshold of fault-tolerant quantum computing~\cite{barends2014superconducting}. The number of qubits in both superconducting and trapped ions quantum computers are both greater than $20$ now \cite{PhysRevX.8.021012,IBMQ}. It is projected that the number of qubits will approach to $50$ or more in the next few years. At that time, the quantum computer may become more powerful than the fastest classical computer for some specific tasks, into the regime of the so called quantum supremacy~\cite{preskill2012quantum}.\\

The IBM Q is a quantum cloud service released by IBM. Its present backend devices include two processors with 5 superconducting qubits ({\it ibmqx2} and {\it ibmqx4}), one 16 qubit processor ({\it ibmqx5}) and one 20 qubit processor ({\it QS1\_1})~\cite{IBMQ}. IBM recently announced that they have successfully built and tested a 20-qubit and a 50-qubit machine~\cite{IBMQ}. The quantum cloud service of IBM provides high fidelity quantum gate operations and measurements. Hence, after the launch of the IBM Q, many groups tested it and performed quantum computing experiments on the cloud; for instance, see~\cite{alsina2016experimental,devitt2016performing,berta2016entropic,Rundle2017,Huffman2017,Hebenstreit2017,2018FA0.22}. \\

Entanglement is considered to be the most nonclassical manifestation of quantum physics~\cite{horodecki2009quantum}. It is also a critical resource for quantum information processing. Highly entangled states such as Bell states, GHZ states and cluster states \cite{briegel2001persistent} have been applied in quantum teleportation, super-dense coding, one-way quantum computing \cite{raussendorf2001one} and various quantum algorithms. The ability to produce highly-entangled states is, therefore, one important step to demonstrate quantumness for quantum processors like {\it ibmqx5}. This task is, however, highly non-trivial due to the error accumulation of faulty gates.

In this paper, we wish to assess the quantumness and performance of the $16$ qubit {\it ibmqx5} device via the production of highly entangled states, namely the graph states, which is an important class of many-body entangled states that are widely used in one-way quantum computing, quantum error correction~\cite{raussendorf2001one,PhysRevLett.98.190504}.
We generate graph states that correspond to rings involving $8$ to $16$ qubits via IBM Q cloud service ({\it ibmqx5}), using optimized low-depth circuits that are tailored to the universal get set on {\it ibmqx5}.
We detect full entanglement up to $16$ qubits, using an entanglement criterion based on reduced density matrices. Qubits are fully entangled in the sense that the state involves all physical qubits and is inseparable with respect to any fixed partition. 

\section{Results}
\label{sec:graphstate}
\subsection{Graph states and entanglement}
\label{sec:EC}
Graph state~\cite{hein2004multiparty} is a generalization of cluster state introduced in 2001~\cite{briegel2001persistent}, which is the resource state of one-way quantum computing~\cite{raussendorf2001one} and quantum error correction~\cite{PhysRevLett.98.190504}.
GHZ state is an example of graph state and has been demonstrated in superconducting qubit system~\cite{Wang2017tenqubit}. However, GHZ state is fragile.
Some other graph states are very robust to local operations, such as local measurements and noises. In order to disentangle the cluster state of $N$ qubits, $~ N/2$ local measurements are needed \cite{briegel2001persistent}.
Because of this nice feature, we decide to generate and detect linear cluster states in the IBM cloud service {\it ibmqx5}.

$X,Y,Z$ denote the Pauli operators. An undirected graph $G\left(V,E\right)$ includes a set of vertices $V$ and a set of edges $E$. A graph state that correspond to an undirected graph $G\left(V,E\right)$ is a $\left|V\right|$-qubit state that has the form
\begin{equation}
\ket{G}=\prod_{\left(a,b\right)\in E}U_{ab}\ket{+}^{\otimes V},
\end{equation}
where $U_{ab}$ is a control-Z operator acting on qubits $a$ and $b$~\cite{Hein2006graph}, and
\begin{equation}
\ket{\pm}=\frac{1}{\sqrt{2}}(\ket{0}\pm\ket{1})
\end{equation}
are eigenvectors of the $X$ operator.

An equivalent definition is, the graph state that corresponds to $G\left(V,E\right)$ is the unique common eigenvector (of eigenvalue $1$) of the set of independent commuting operators:
\begin{equation}
\label{eq:stab}
K_a=X^aZ^{N_a}=X^a\prod_{b\in N_a}Z^b,
\end{equation}
where the eigenvalues to $K_a$ are $+1$ for all $a\in V$, and $N_a$ denotes the set of neighbor vertices of $a$ in $G$~\cite{Hein2006graph}.
 As implied by the first definition, a $n$-qubit graph state can be prepared by the following steps.\\
\indent 1. Initialize the state to $\ket{+}^{\otimes n}$ by applying $n$ Hadamard gates to $\ket{0}^{\otimes n}$;\\
\indent 2. For every $\left(a,b\right)\in E$, apply a control-Z gate on qubits $a$ and $b$; the order can be arbitrary.\\

Entanglement of general mixed states was discussed by Werner in 1989~\cite{werner1989quantum}. Since then, many entanglement criteria were proposed; among them the widely used ones include the partial transpose criterion~\cite{peres1996separability,Horodecki1996ppt,horodecki2009quantum} and the symmetric extension criterion~\cite{doherty2002distinguishing}.

A bipartite state $\rho_{AB}$ on the Hilbert space $\mathcal{H}=\mathcal{H}_A\otimes\mathcal{H}_B$ is said to be separable if $\rho_{AB}$ can be written as
\begin{equation}
\rho_{AB}=\sum_i p_i\rho_A^{i}\otimes\rho_B^{i},
\end{equation}
where $\rho_A^{i}$ and $\rho_B^{i}$ are quantum states of the system $A$ and $B$, respectively, with $p_i\geq 0$ and $\sum_i p_i=1$. Otherwise $\rho_{AB}$ is entangled. For a state $\rho$ of a many-body system, for any fixed bipartition $AB$ of the system, if $\rho$ is entangled with respect to the partition $AB$, then the entanglement of the many-body state $\rho$ can also be examined via its subsystems. That is, if the subsystems are all entangled, the whole system must be also entangled.

To be more concrete, consider a 4-qubit subsystem $\rho_{A,B,C,D}$ in an $n$ qubit system. Suppose that we perform two local operations $O_A$ and $O_D$ on qubit $A$ and $D$ respectively, and then obtain the reduced density matrix of qubit $B$ and $C$ by tracing out qubit $A$ and $D$. The reduced density matrix for qubits $B$ and $C$ reads
\begin{equation}
\rho'_{B,C}=tr_{A,D}\left(\frac{O_AO_D\rho_{A,B,C,D}O_D^{\dagger}O_A^{\dagger}}{tr\left(O_AO_D\rho_{A,B,C,D}O_D^{\dagger}O_A^{\dagger}\right)}\right).
\end{equation}

The entanglement of $\rho'_{B,C}$ can be determined by using entanglement monotones such as negativity and concurrence, which, in the 2 qubit case, has non-zero values if and only if the system is entangled~\cite{Horodecki1996ppt,horodecki2009quantum}. If $\rho'_{B,C}$ is entangled, we can conclude that in the original system, there could not exist a separation with qubit $B$ and $C$ on different sides. In other words, if the original system is biseparable with respect to a fixed partition, the qubit $B$ and $C$ must be on the same side. Otherwise, we will be able to create entanglement between the two separable parties with only local operations, which is not possible~\cite{horodecki2009quantum}.

For an $n$ qubit system $\{q_1,q_2,...,q_n\}$, if we can show that among the $n$ qubit pairs $\left(q_1,q_2\right)$,...,$\left(q_{n-1},q_n\right)$,$\left(q_n,q_1\right)$, $n-1$ of them must be on the same side in a separation, then we may conclude that there is no possible separation, and that the system is a $n$-qubit entangled state (meaning that the state is not biseparable w.r.t. a fixed partition, and that it involves all qubits). The (minimal) number of circuit configurations needed in this approach is $3^4\left(n-1\right)$, which grows linear with respect to $n$. This method is far more efficient compared to a full $n$-qubit tomography, which requires exponential number of configurations.\\

\subsection{Graph states on ibmqx5}
\label{sec:ibmqx5}
\begin{figure}[H]
  \centering
  \includegraphics[width=0.8\columnwidth]{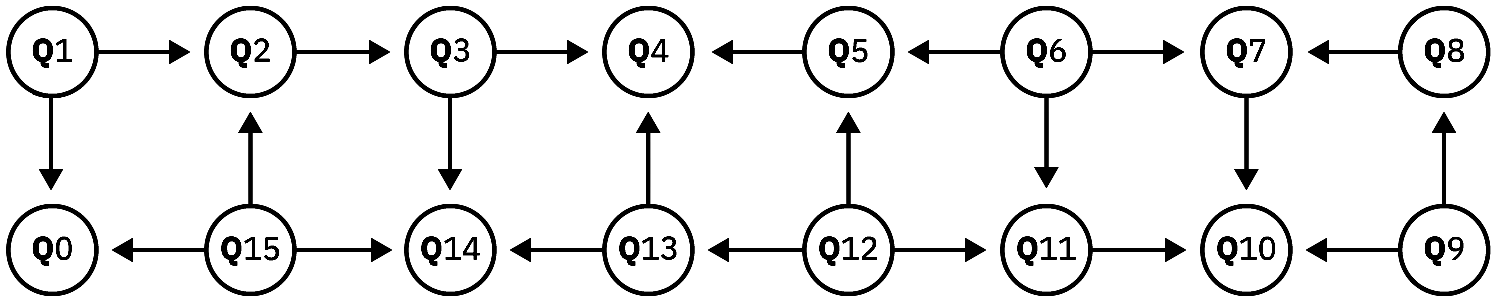}
    \caption{Connectivity map of {\it ibmqx5}~\cite{ibmqx5}}
 \label{fig:topo1}
\end{figure}

\begin{figure}[htpb]
	\centering
	\includegraphics[width=0.95\columnwidth]{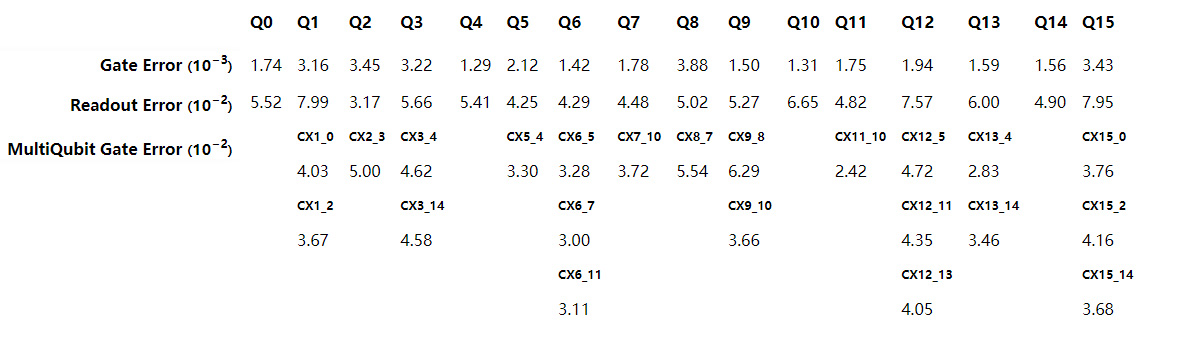}
	\caption{Calibration parameters of {\it ibmqx5}, archived 2018-01-10 from \cite{IBMQ}. It should be noted that these parameters are updated on a daily basis.}
	\label{fig:parameter}
\end{figure}

{\it ibmqx5} is a 16-qubit superconducting quantum processor. It allows independent single qubit operations with fidelity $>99\%$ and control operations with fidelity $95-97\%$ (see Fig. \ref{fig:parameter}) 
marked as the edges in the connectivity map (see Fig.\ref{fig:topo1}). That is, CNOT operations with qubit $a$ as the control qubit and $b$ as the target is allowed if and only if $a\rightarrow b$ is an edge in the map.

\begin{figure}[htpb]
  \centering
  \includegraphics[width=0.8\columnwidth]{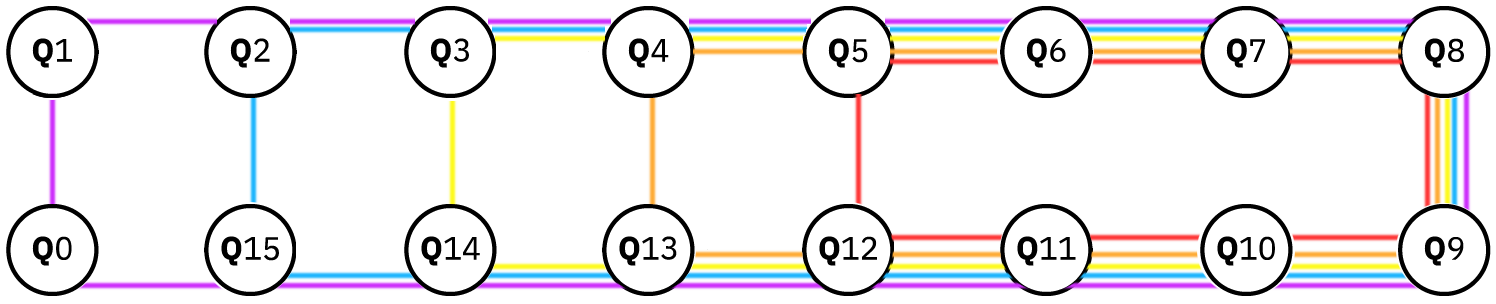}
  \caption{Graph states employed in this experiment. Colored lines illustrate the graph of the 8 qubit graph state(in red), 10 qubit graph state(orange), 12 qubit graph state(yellow), 14 qubit state(blue) and 16 qubit graph state(purple).}
 \label{fig:state1}
\end{figure}

\begin{figure}[htpb]
\centering
	\subfigure{
		\label{fig:circuit2}
	}
	\subfigure{
		\label{fig:circuit3}
	}
\includegraphics[height=0.2\textheight]{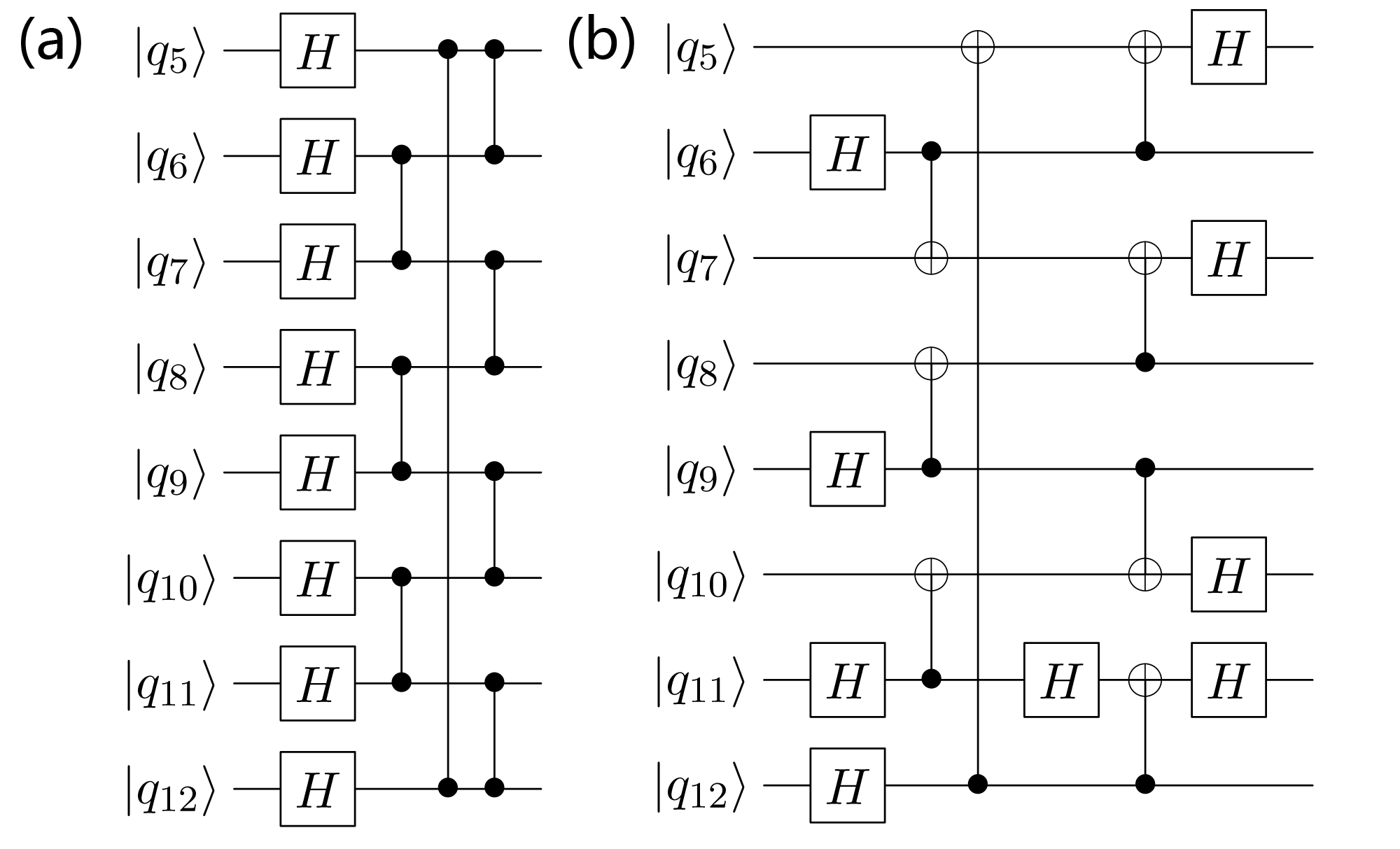}
 \caption{(a) The quantum circuit for preparing a $8$-qubit graph state implied by the definition of graph states; (b) The optimized circuit that suits {\it ibmqx5}'s connectivity.}
 \label{fig:circuit1}
\end{figure}

In our experiment, as shown in Fig. \ref{fig:state1}, the following five graph states are employed. The first state is a 8 qubit graph state involving qubits $q5-q12$ that corresponds to a ring of length 8; the second one is a 10 qubit state involving qubits $q4-q13$ corresponding to a ring of length 10; the third one
involves qubits $q3-q14$ and corresponds to a ring of length 12; the 4th one involves qubits $q2-q15$ and corresponds to a ring of length 14; the 5th one involves all the 16
qubits.
We employ these particular graph states based on the following considerations. First, these states are genuinely entangled and will remain entangled after tracing out a large number of qubits. Second, research has shown that 1-D cluster states are robust against decoherence, meaning that it would be more likely to find entanglement in a rather large graph state close to a 1-D chain, compared to GHZ states and 2-D graph states~\cite{hdb2016decoherence}. At last, even rings are two-edge-colorable; as a result, on the 16-qubit ibmqx5, these ``even-ring'' states could be prepared using low-depth circuits (See Fig.~\ref{fig:circuit1}).

\begin{figure}[htpt]
\centering
  \subfigure{
  \label{fig:circuit4}
  }
  \subfigure{
  \label{fig:circuit5}
  }\\
    \subfigure{
  \label{fig:circuit6}
  }
  \subfigure{
  \label{fig:circuit7}
  }
\includegraphics[height=0.6\textheight]{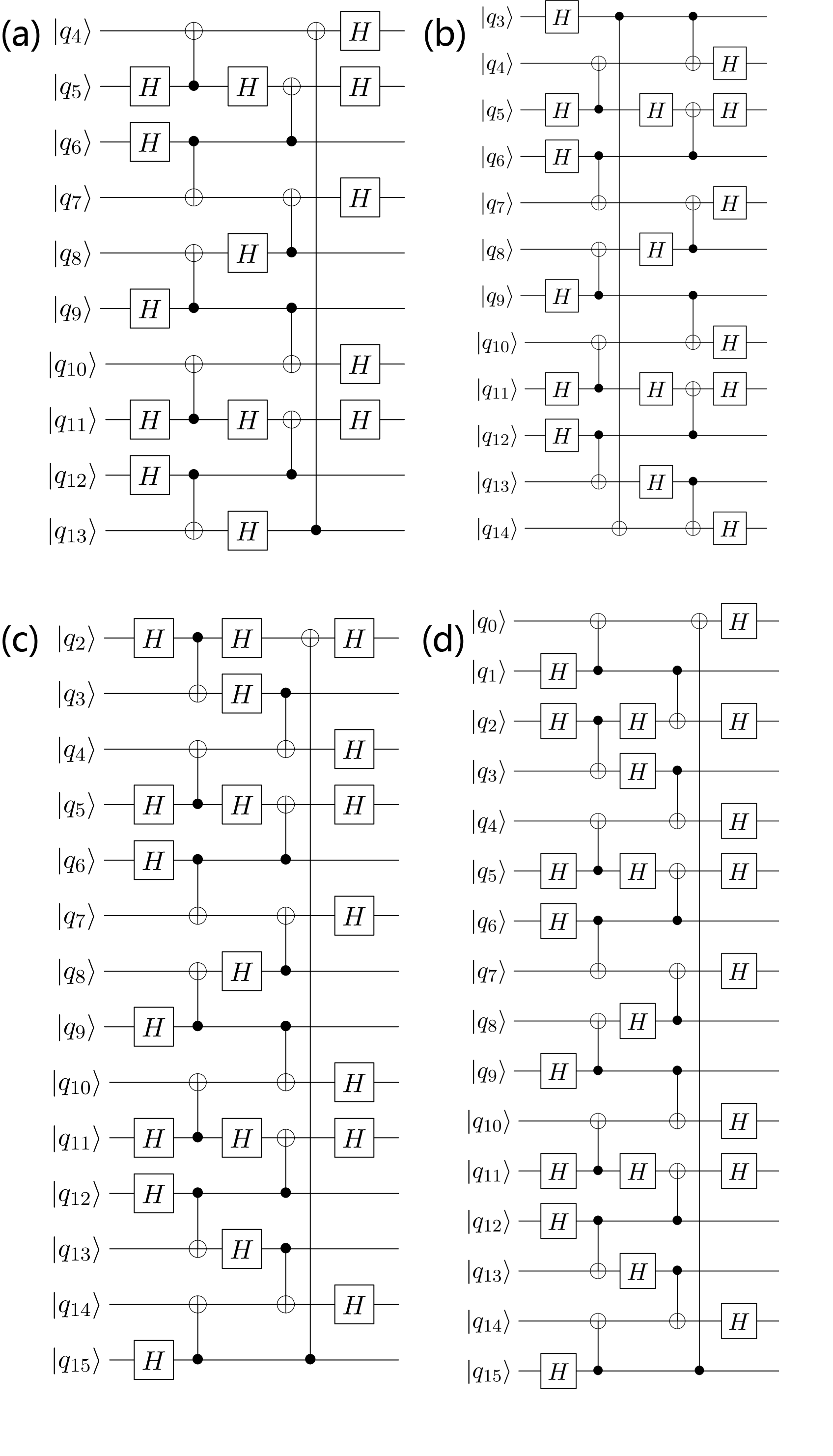}
  \caption{The quantum circuit implemented on {\it ibmqx5} for the preparation of (a)$10$ qubit graph state; (b)$12$ qubit graph state; (c)$14$ qubit graph state; (d)$16$ qubit graph state.}
\end{figure}

To prepare the desired graph state, we start from the circuit implied by the definition of graph states (see Fig.~\ref{fig:circuit2}). The control-Z gates are implemented using a CNOT gate and two Hadamard gates. We then optimize this circuit by adjusting the order of commuting gates and removing redundant Hadamard gates (see Fig.~\ref{fig:circuit3}). The circuit that we implemented are shown in Fig.~\ref{fig:circuit3} and Fig.~\ref{fig:circuit4}-Fig.~\ref{fig:circuit7}.

\subsection{Experimental Results}
\label{sec:result}
For each $n$-qubit ring state, $n$ partial tomographies are performed for every subsystem with $4$ qubits that forms a chain in the ring. E.g., for the $8$-qubit graph state, the $8$ subsystems are $\left(q5,q6,q7,q8\right)$, $\left(q6,q7,q8,q9\right)$,..., $\left(q12,q5,q6,q7\right)$. For every state, $3^4n$ experimental configurations are used; $2048$ measurements are taken under each configuration. The $n$ $4$-qubit reduced density matrices are obtained using the maximum likelihood method proposed by J.Smolin et. al~\cite{Smolin2012mle}.\\
\begin{figure}[hbt]
  \centering
  \includegraphics[width=0.8\columnwidth]{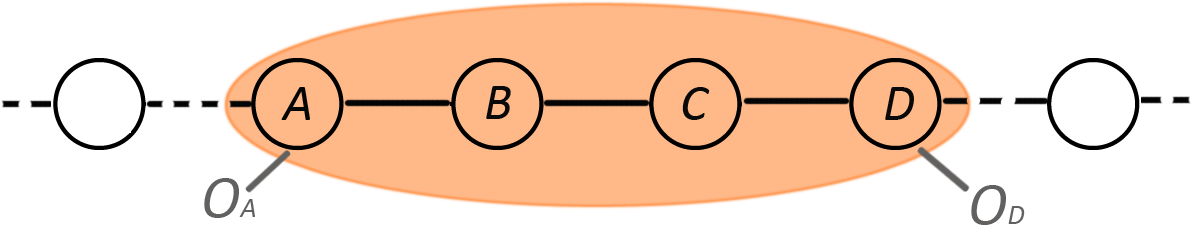}
  \caption{A four-qubit subsystem that forms a chain}
 \label{fig:subsytem1}
\end{figure}

Due to Eq.~\eqref{eq:stab}, for a ring graph state, each $4$-qubit density matrix of neighboring four qubits, as illustrated in Fig~\ref{fig:subsytem1} is given by
\begin{equation}
\rho_{A,B,C,D}=\frac{1}{4}(I+Z_AX_BZ_C)(I+Z_BX_CZ_D).
\end{equation}

\begin{figure}[htpt]
	\centering
	\subfigure{
		\label{fig:result8}
	}
	\subfigure{
		\label{fig:result10}
	}
	\subfigure{
		\label{fig:result12}
	}
	\subfigure{
		\label{fig:result14}
	}
	\subfigure{
		\label{fig:result16}
	}
	\includegraphics[width=\textwidth]{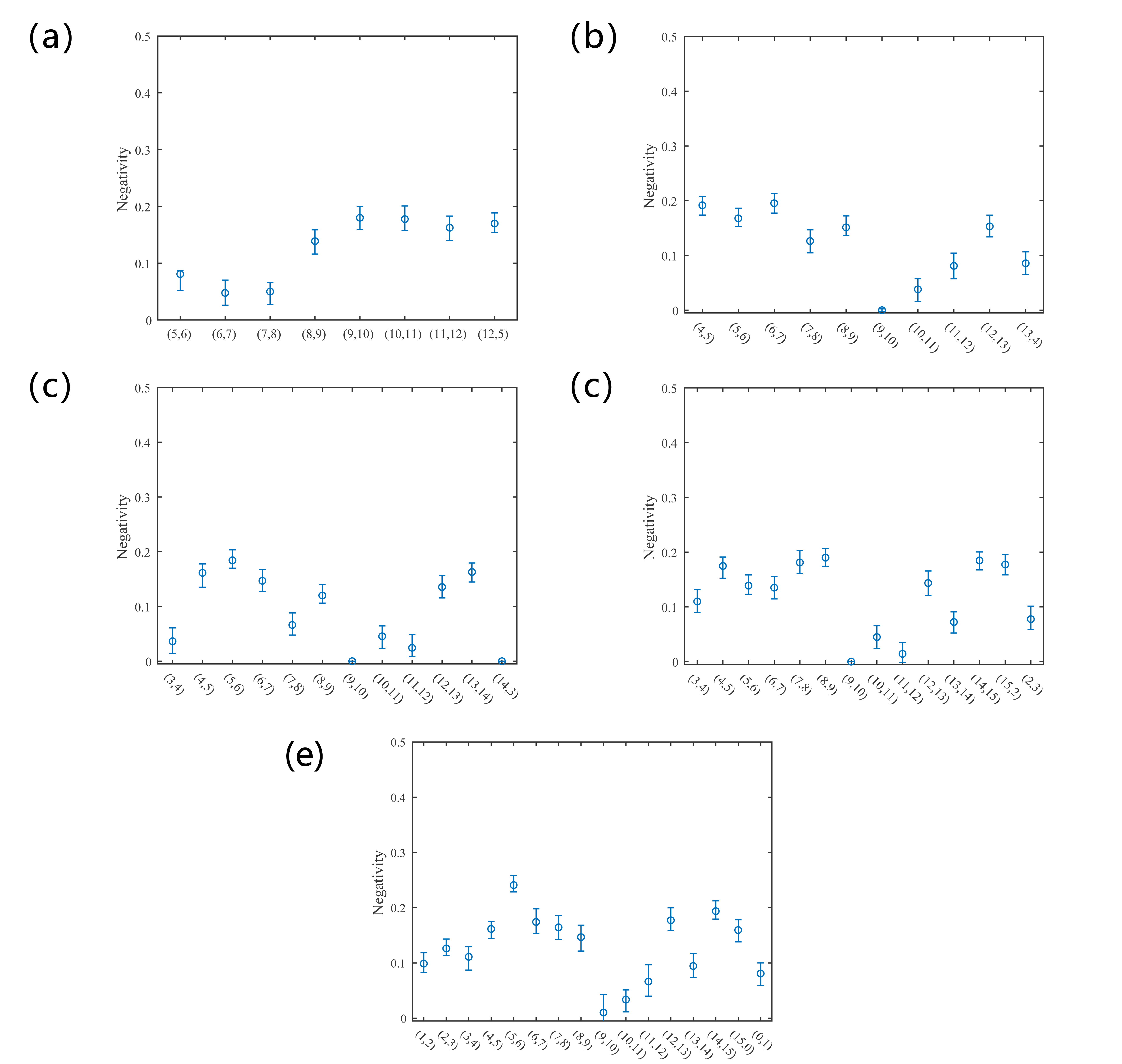}
	\caption{The result of (a) the $8$ qubit graph state; (b) the $10$ qubit graph state; (c) the $12$ qubit graph state; (d) the $14$ qubit graph state; (e) the $16$ qubit graph state. The negativity of the final $2$-qubit states are plotted. $95\%$ confidence intervals are estimated using bootstrapping techniques.}
	\label{fig:allresult}
\end{figure}

Then, for each $4$-qubit density matrix, we apply the local operations $O_A=\frac{Z_A+I}{2}$ and $O_D=\frac{Z_D+I}{2}$ and calculate the negativity of the resulting 2-qubit subsystem. For instance, we may choose $\left(q5,q6,q7,q8\right)$ as our subsystem; after applying $O_A$ and $O_D$ to $q5$ and $q8$ respectively, we will trace out $q5$ and $q8$, and measure the negativity of the remaining subsystem, $\left(q6,q7\right)$.
We choose $O_A=\frac{Z_A+I}{2}$ and $O_D=\frac{Z_D+I}{2}$ for the following reason. If $\rho$ is graph state, and the $4$-qubit subsystem corresponds to $4$ vertices that form a chain in the graph, then the resulting $2$-qubit state is a maximally entangled state
\begin{equation}
\ket{\phi}=\frac{1}{\sqrt{2}}\left(\ket{0}\ket{+}+\ket{1}\ket{-}\right).
\end{equation}
Therefore, for a state close to this graph state, we should expect the resulting $2$-qubit state to have a negativity significantly greater than zero.
The results are plotted in Fig.\ref{fig:allresult}. 

For the $8$-qubit graph state, the measured negativities are all significantly greater than $0$. For the $10$-qubit graph state, $9$ out of $10$ measured negativities are significantly greater than $0$. Based on our argument in Sec. \ref{sec:EC}, both the $8$-qubit state and $10$-qubit state are fully entangled.

 In the $12$ qubit case, as shown in Fig. \ref{fig:result12}, $10$ out of $12$ measured negativites are significantly non-zero. The two zeros
come from $(q9,q10)$ and $(q14,q3)$ pairs.
 Therefore, there is only one possible separation, namely $\{q10,q11,q12,q13,q14\}\left.\right|\{q3,q4,q5,q6,q7,q8,q9\}$. Should this be true, the reduced density matrix of qubits $q8,q9,q10,q11$ should also be separable with the separation $\{q8,q9\}\left.\right|\{q10,q11\}$. In that case, its partial transpose with respect to qubit $q8$ and $q9$ must be positive. However, with respect to this partial transpose, $\rho_{q8,q9,q10,q11}$ has negativity $0.0391\pm 0.0039$ (standard deviation estimated via bootstrapping). Therefore, this possibility is ruled out with very high confidence. We can now conclude that the $12$ qubit graph state is fully entangled.

In the $14$ qubit case, as shown in Fig. \ref{fig:result14}, $12$ out of $14$ measured negativites are significantly greater than 0. Here, we may apply the same trick again. The only possible separation is $\{q2,q3,q4,q5,q6,q7,q8,q9,q12,q13,q14\}\left.\right|\{q10,q11\}$. In this case, subsystem $\{q8,q9,q10,q11\}$ should have zero negativity with respect to the partial transpose on $q8$ and $q9$. However, the measured negativity is $0.0698\pm 0.0048$(standard deviation estimated via bootstrapping). Hence, this possibility is ruled out with very high confidence. We may conclude that this state is fully entangled.

In the $16$ qubit case, as shown in Fig. \ref{fig:result16}, $15$ out of $16$ measured negativites are significantly greater than 0. As argued in Sec. \ref{sec:EC}, this means that this state is not biseparable w.r.t. a fixed partition, thereby showing that all 16 qubits in {\it ibmqx5} are in full entanglement.

It may be noted that the subsystem of qubits $\{q8,q9,q10,q11\}$ yields close-to-zero negativity in 3 out of 4 experiments. This can be due to relatively high readout errors or gate errors involving these qubits, which is compatible with the measured parameters provided by IBM's website~\cite{IBMQ} (see Fig. \ref{fig:parameter}). For instance, the CNOT gate between $q10$ and $q11$ has the largest error among all possible CNOT gates, while the readout error of $q10$ and $q11$ are also above the average level ($6.5\%$).

\subsection{Further Exploration of the 16-qubit State}
\label{sec:further}
The results above could be understood as an ability to generate localized entanglement on physically neighboring qubits \cite{LE}. That is, neighboring qubits can be put into entanglement by performing ideal local operations on the 16-qubit state. Using the same data obtained above, we will show that localized entanglement on qubits with distance 2 and 3 could also be generated.

\begin{figure}[htpb]
  \centering
  \includegraphics[width=0.8\columnwidth]{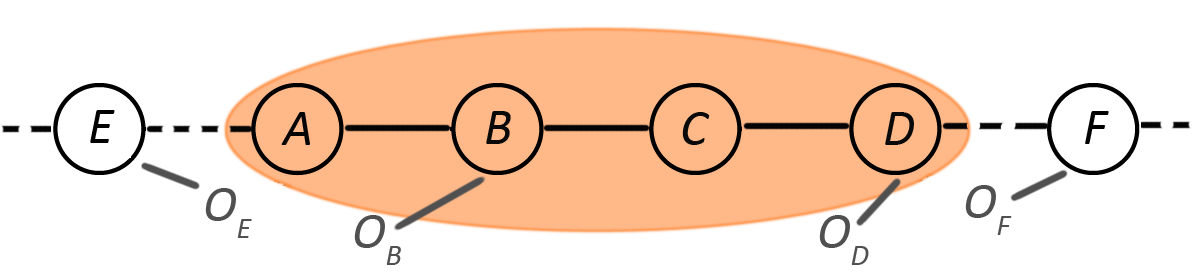}
  \caption{Operations performed to produce entanglement on subsystem $\{A,C\}$.}
 \label{fig:operationbd}
\end{figure}

Suppose $\{E,A,B,C,D,F\}$ is a six-qubit subsystem that forms a chain. We first apply $O_E=\frac{Z_E\pm I}{2}$ and $O_F=\frac{Z_F\pm I}{2}$ on $E$ and $F$ respectively(four possibilities). On our data, this can effectively be done by first postselecting 0s on qubits $E$ and $F$ before calculating the tomography of $\{A,B,C,D\}$. Next, $O_B=\frac{X_B+I}{2}$ and $O_D=\frac{Z_D+I}{2}$ are performed (see Fig.\ref{fig:operationbd}). At last, $B$, $E$, $F$ and $D$ are traced out, while the negativity in subsystem $\{A,C\}$ is calculated. If the 16-qubit state is perfect, this resulting system would be maximum entangled.

\begin{table}[H]\footnotesize
	\centering
	\caption{Negativities of qubits with distance 2 in the $16$-qubit state}
	\begin{tabular}{|c|c|c|c|c|c|c|c|}
		\hline {(0,2)}&(1,3)&(2,4)&(3,5)&(4,6)&(5,7)&(6,8)&(7,9)\\
		\hline 0.023&0.027&0.088&0.145&0.143&0.156&0.134&0.105\\
		\hline
		{(8,10)}&(9,11)&(10,12)&(11,13)&(12,14)&(13,15)&(14,0)&(15,1)\\
		\hline
		0.178&0&0.114&0.079&0.040&0.028&0&0\\
		\hline
	\end{tabular}
	\label{tab:neg1}
\end{table}

Based on data obtained in previous experiments, we have calculated the corresponding negativity for each 6-qubit subsystem and shown them in Table.~\ref{tab:neg1}. Using this method, we have identified localized entanglement in 13 out of 16 pairs of qubits with distance 2.

\begin{figure}[htpb]
  \centering
  \includegraphics[width=0.8\columnwidth]{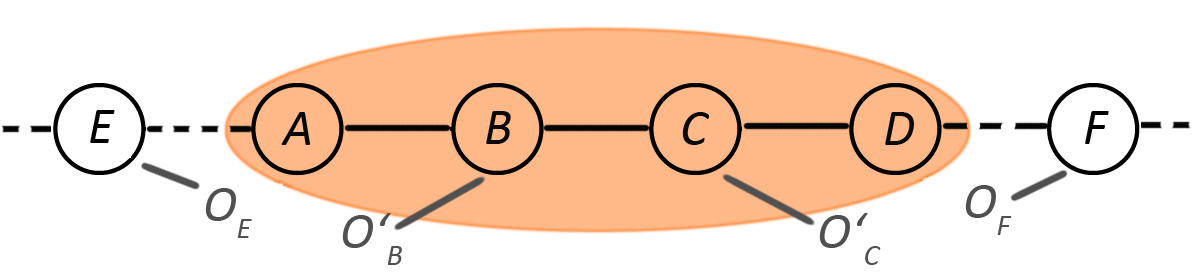}
  \caption{Operations performed to produce entanglement on subsystem $\{A,D\}$.}
 \label{fig:operationad}
\end{figure}
To generate localized entanglement on qubits with distance 3, we may apply the same $O_E$ and $O_F$, and then apply $O'_B=\frac{X_B+I}{2}$ and $O'_C=\frac{X_C+I}{2}$ (see Fig.\ref{fig:operationad}). The negativity of subsystem $\{A,D\}$ would be calculated. Again, if the 16 qubit state is perfect, this two qubits would be maximum entangled; therefore, we should expect a non-zero negativity if the actual state is close to the theoretical one.

Among 16 pairs of qubits with distance 3, we have identified localized entanglement in 6 pairs of them. The results based on our data is presented in Table.~\ref{tab:neg2}.
 \begin{table}[htbl]\footnotesize
	\centering
	\caption{Negativities of qubits with distance 3 in the $16$-qubit state}
	\begin{tabular}{|c|c|c|c|c|c|c|c|}
		\hline {(0,3)}&(1,4)&(2,5)&(3,6)&(4,7)&(5,8)&(6,9)&(7,10)\\
        \hline 0&0&0.085&0.093&0.110&0.097&0.061&0\\
        \hline
        {(8,11)}&(9,12)&(10,13)&(11,14)&(12,15)&(13,0)&(14,1)&(15,2)\\
        \hline
        0.012&0&0&0&0&0&0&0\\
        \hline
	\end{tabular}
	\label{tab:neg2}
\end{table}

\section{Discussion}
\label{sec:final}
We have prepared graph states of $8$, $10$, $12$, $14$ and $16$ qubits on the 16-qubit {\it ibmqx5} processor and demonstrated that these graph states are not biseparable w.r.t. any fixed partition. In particular, we have realized full entanglement using all 16 qubits. Moreover, we have demonstrated the ability to create localized entanglement on qubit pairs with distance 3 and 4 from this 16-qubit entangled state. In our approach of detecting nonseparability, we only have to measure the reduced density matrix of up to $4$ qubits, and the size of reduced density matrix does not scale with the total qubit number, i.e.~our method is efficient and scalable. In our experiments, graph states do not have high fidelity because of the large number of qubits, e.g.~the fidelity of the $12$-qubit graph state is lower than $0.44$. (This upperbound is obtained by computing the fidelity between each 4-qubit subsystem and the theoretical result and taking the minimum.) However, the negativity of $4$-qubit reduced density matrix decays gently with respect to the qubit number, which implies that the error per qubit weakly depends on the qubit number. It is a strong evidence that {\it ibmqx5} is capable of generating highly entangled states and demonstrates the computer's quantumness. In computational tasks such as one-way quantum computing, graph state with decaying fidelity is acceptable, and the computing is fault-tolerant as long as the error per qubit is lower than a threshold~\cite{Nielson2005FaultTolerant,Raussendorf2006FaultTolerant}.

\section*{Data Availability}
The experimental data that support the findings of this study\cite{DataSet} are available in figshare with the identifier \href{http://dx.doi.org/10.6084/m9.figshare.6790781}{10.6084/m9.figshare.6790781}.

\section*{Acknowledgments}
We gratefully acknowledge the IBM-Q team for providing us with access to their 16-qubit platform.
The views expressed are those of the authors and do not reflect the official policy or position of IBM or the IBM Quantum Experience team.

\section*{Competing Interests}
The authors declare no competing interests..

\section*{Author Contributions}
Y.L., Z.Y. and B.Z. designed and conceived the study. Y.W. designed quantum circuits and performed the experiments.
All authors wrote the manuscript.

\section*{Funding}
Y.L. is supported by NSAF (Grant No. U1730449).
Z.-Q.Y. is supported by the National Natural Science Foundation of China Grant 61771278, 11574176 and 11474177. B.Z. is supported by NSERC and CIFAR.


\end{document}